\documentclass[preprint,showpacs,prb]{revtex4}

\usepackage{graphicx}
\usepackage{dcolumn}
\usepackage{bm}
\usepackage{color}

\begin{document}

\preprint{Applied Physics Letters \textbf{91}, 041114(2007).}

\title{High speed single photon detection in the near-infrared}

\author{Z. L. Yuan}
\email{zhiliang.yuan@crl.toshiba.co.uk}

\author {B. E. Kardynal}

\author {A. W. Sharpe}

\author {A. J. Shields}

\affiliation{Toshiba Research Europe Ltd, Cambridge Research
Laboratory, 208 Cambridge Science Park, Milton Road, Cambridge, CB4
0GZ, UK }

\date{\today}

\begin{abstract}
InGaAs avalanche photodiodes (APDs) are convenient for single photon
detection in the near-infrared (NIR) including the fibre
communication bands (1.31/1.55~$\mu$m).  However, to suppress
afterpulse noise due to trapped avalanche charge, they must be gated
with MHz repetition frequencies, thereby severely limiting the count
rate in NIR applications. Here we show gating frequencies for
InGaAs-APDs well beyond 1 GHz. Using a self-differencing technique
to sense much weaker avalanches, we reduce drastically afterpulse
noise. At 1.25~GHz, we obtain a detection efficiency of 10.8\% with
an afterpulse probability of 6.16\%. In addition, the detector
features low jitter (55~ps) and a count rate of 100~MHz.
\end{abstract}


\maketitle

Near-infrared (NIR) single photon detection is required for quantum
key distribution (QKD),\cite{bennett84} fibre optical
sensing,\cite{opticalsensing} eye-safe ranging and
timing,\cite{maruyama02} semiconductor device
analysis,\cite{ispasoiu06} singlet oxygen detection,\cite{shimizu97}
as well as photonic research in a wide range of disciplines.
InGaAs/InP avalanche photodiodes (APDs) have proven a practical
choice for wavelengths in the region
1.0--1.7~$\mu$m.\cite{lacaita96,tomita02,bethune04,yoshizawa04,namekata06}
However, the performance of these devices lags far behind that of
silicon APDs, used for visible photon detection, due to greater
defect density in III-V semiconductor devices.

An APD detects a single photon through avalanche multiplication of a
photon-excited carrier, resulting in a detectable macroscopic
current flow. However, avalanche carriers that are trapped by
defects, and then subsequently spontaneously released, can induce a
second spurious avalanche, contributing to the erroneous count
noise. To suppress the rate of these "afterpulses", InGaAs APDs are
normally gated with only MHz frequencies with each active gate
lasting for a few nanoseconds. Furthermore, a dead time of a few
microseconds,\cite{bethune04} has to be applied following each
photon detection to allow the trapped carriers time for decay. These
considerations limit the maximum count rate of an APD to typically
just $\sim$100~kHz.\cite{yoshizawa04} This has rendered InGaAs APDs
unsuitable for applications, which require high-speed single photon
detection, such as next generation high bit rate
QKD.\cite{gordon05,thew06,diamanti06}

Alternative technologies, such as upconversion with periodically
poled LiNbO$_3$ (PPLN)\cite{albota04} and superconducting nanowire
detectors,\cite{hadfield06} have been developed. However, those PPLN
upconversion detectors\cite{thew06,diamanti06} suffer from high
background photon count rate and require sophisticated optical
alignment for noise rejection, while superconducting device requires
cryogenic cooling to temperatures of around a few K.

In this paper, we show gating frequencies for InGaAs-APDs well
beyond 1~GHz. Using a self-differencing technique to sense much
weaker avalanches, we reduce drastically the avalanche charge and
hence the afterpulse noise. At 1.25~GHz, we obtain a detection
efficiency of 10.8\% with an afterpulse probability of 6.16\%. In
addition, the detector features low timing jitter (55~ps) and a
count rate of 100~MHz, and does not require cryogenic cooling. These
features make the device suitable for a wide range of applications
requiring high count rate, low noise, fast NIR single photon
detection.

The key to improve InGaAs APD speed is to limit the avalanche
charges by detecting weak avalanches. Unfortunately, weak avalanche
signals are often buried within the APD capacitive response,
especially when using fast gating signals. Figure 1(b) plots an APD
response, recorded on an oscilloscope with 100-ps rise time
resolution, to a series of 0.62~GHz square wave voltage gating
pulses (Fig.~1(a)). In response to each gate, the APD produces a
positive peak, followed by a negative peak, due to the charging and
subsequent discharging of the finite capacitance of the APD. No
avalanche signals can be identified in this waveform. However,
numerically subtracting a signal that is identical to Fig.~1(b) but
shifted by one gating clock as shown in Fig.~1(c), the resulting
waveform (Fig.~1(d)) shows clearly a distinctive positive peak,
followed by a negative peak separated by one clock period. The
positive peak and the following negative one revealed here is
attributed to an avalanche. By discriminating either the positive or
negative peaks, avalanches, previously completely buried within the
capacitive response, become detectable. Note that the amplitude of
the avalanche signal is 10 times weaker than the APD capacitive
response and is superimposed on the negative capacitive peak.
Without removing the capacitive response, an avalanche would have to
be at least 20 times stronger in order to be detected.

Numerical subtraction is rather impractical, especially when the
gating frequency is high. In practice, such an operation may be
realized using hardware. Figure~1(e) shows a self-differencing
circuit designed for operation that is equivalent to the numerical
subtraction described above. It consists of a 50/50 pulse splitter
to divide the APD response into two equal components, which are then
combined by a differencer. The coaxial cables connecting the
splitter and the differencer have different lengths to induce a
delay of one gating period between two components. Thus, the output
of the self-differencer is the difference between two identical
waveforms shifted relatively by one gating period. A 21~dB
suppression of the APD capacitive response is achieved with the
circuit. Figure 1(f) shows a detectable avalanche as a result of
self-differencing.

The InGaAs APD under test was cooled electrically to $-30^\circ$C.
It was gated using a square wave signal generated by a pulse
generator. The APD output, after passing through the
self-differencer, is pulse-shaped using a discriminator before
analysis by a pulse counter or a time-correlated photon counter. To
characterize its detection efficiency, the APD was illuminated with
80-ps long pulses of 1550~nm light from a distributed feedback laser
diode, which was synchronized at 1/64 of the APD gating frequency.
Each pulse was attenuated to 0.1~photons per pulse before coupling
into the APD fibre pigtail. The setup is suitable for measuring the
detection efficiency, dark count and afterpulse probabilities.

We first tested whether the InGaAs APD detects single photons by
scanning the laser pulse delay. The APD was biased at 1.4~V below
its breakdown voltage of 47.3~V, and gated by a superimposed
0.62~GHz square wave with an amplitude of 6.6~V. As shown in
Fig.~2(a), where the photon count rate is plotted as a function of
the laser pulse delay, photons are detected only when they arrive
within the gate duration defined by the applied square wave. Photons
arriving between the gates are not detected, evidenced by the fact
that the count rate falls to the dark count level. The full width at
half maximum (FWHM) of the photon detection peak is 170~ps,
significantly narrower than the gate width (800~ps), suggesting that
the APD active duration is much smaller than the electrical gate.

Figure~2(b) shows a time-resolved histogram of photon arrivals when
the laser delay was tuned to maximize count rate. As the laser diode
was triggered at the 1/64 of the gating frequency, the peak at 0~ns
corresponding to the illuminated gate is much stronger than the
remaining peaks corresponding to the non-illuminated gates. The
photon-induced peak is very sharp, with a FWHM of 55~ps, indicating
extremely low detection jitter. Adjacent peaks are well separated
with a $>$700~ps gap within which no counts were registered.

The histogram of dark counts is shown in Fig.~2(c), measured with
the laser switched off. The count rate is much lower for each gate
than recorded for non-illuminated gates (Fig.~2(b)) with the laser
on, suggesting that under illumination afterpulses are the dominant
noise source. The afterpulse probability $P_{A}$, defined here as
the ratio of the total afterpulse counts to the photon counts, can
be obtained from
\begin{equation}
P_{A}=\frac{(I_{NI}-I_D)\cdot R}{I_{Ph}-I_{NI}}
\end{equation}
where $I_{Ph}$ and $I_{NI}$ are the count rate per gate at the
illuminated and non-illuminated gates respectively, while $I_D$ is
the dark count rate for each gate. $R=64$ is the ratio of the gating
frequency to the laser pulse frequency.

The afterpulse probability was measured as a function of the photon
detection efficiency, which was varied by tuning the DC bias level
in the range of 45.5 --- 47.0~V. The square wave amplitude was fixed
at 6.6~V for a gating frequency of 0.62 or 0.98~GHz, and at 4.6~V
for 1.25~GHz. In Fig.~3(a), $P_{A}$ is plotted as a function of the
net detection efficiency $\eta$ (excluding the dark and afterpulse
counts). $P_{A}$ generally increases with $\eta$, consistent with
the greater avalanche charge flow at the higher bias required for
higher detection efficiencies. There appears to be a critical
efficiency $\eta_{c}$, below which $P_A$ is less than 6.2\% and
changes slowly with $\eta$. When $\eta>\eta_{c}$, $P_{A}$ increases
sharply. It is found that $\eta_{c}\doteq21\%$ for 0.62~GHz and
becomes notably lower at $\sim$11\% for 0.98~GHz and 1.25~GHz. It
could well suggest that afterpulse noise is more prominent at higher
gating frequencies. However, this may not be the case because $\eta$
may be underestimated for high frequencies, due to the comparable
duration of the optical active gate and the laser pulse duration.

Figure~3(b) plots the dark count probability $P_D$ as a function of
$\eta$. As usual, $P_D$ increases with $\eta$. At $\eta\approx11\%$,
the dark count probability is similar for all gating frequencies, at
$\sim2.5\times10^{-6}$ per gate. Such a dark count probability is
very low indeed, considering the operation temperature of
-30$^\circ$C. It is roughly 20 times less than obtained previously
using conventional discrimination.\cite{yuan07}

The detector linearity and maximum achievable count rate were also
studied. In this experiment, a standard telecom direct modulation
laser diode was used to illuminate the APD at the same rate as the
APD gating. The relatively wide optical pulse (250~ps FWHM) resulted
in an effectively lower photon detection efficiency.  As shown in
Fig.~4, the count rate increases linearly with the photon flux over
$>30$~dB dynamical range. The increase in photon count rate is
sub-linear for count rates exceeding 20~MHz, and finally saturates
at 100~MHz. The count rate here is more than two orders of magnitude
higher than previously achieved.\cite{yoshizawa04} Moreover, the
saturation rate is limited by the pulse-shaping electronics, rather
than the detector, and could be increased further by using faster
electronics.

\begin{table} [t]
\caption{\label{tab:table1}Performance comparison of the
self-differencing InGaAs APD with the PPLN upconversion detectors.\\
}
\begin{ruledtabular}
\begin{tabular}{lcccc}
  & $\eta$ & $P_D$ & $\eta/P_D$ & Max (MHz) \\
\hline InGaAs APD & $10.9\%$&$2.34\times10^{-6}$&$4.6\times10^4$
&100
\\
PPLN\footnotemark[1]&1.2\% & $7\times10^{-6}$ & $1.71\times10^3$ &15
\\ 
PPLN\footnotemark[2]& 3.6\% &$1.95\times10^{-5}$&$1.85\times10^3$
&15
\\
\end{tabular}
\end{ruledtabular}
\footnotetext[1] {Reference 12} \footnotetext[2] {Reference 13}
\end{table}

Finally we compare the performance of the self-differencing APD with
PPLN up-conversion detectors using Si
detectors.\cite{thew06,diamanti06} As summarized in Table I, the
self-differencing APD shows significantly better detection
efficiency and dark count probability $P_D$. It is also polarization
independent and sensitive to a broad spectrum of photons while an
up-conversion detector only detects photons at a fixed wavelength
with a particular polarization due to stringent phase matching
requirement. Moreover, the self-differencing APD features $100$~MHz
maximum count rate, as compared with $\sim$15 MHz for an
up-conversion detector,\cite{perkinelmer} making it more attractive
for high bit rate QKD. Besides improved performance, gated operation
is another advantage for QKD. Gated operation automatically rejects
background photons arriving between gates. As demonstrated in
Fig.~2(a) the APD active width is only 170 ps, far less than the
gating period of 1.6~ns, suggesting $\sim90\%$ rejection of
background photons. Of course, gated operation requires clock
synchronization, however, even more precise synchronization is
required for the up-conversion detectors based QKD where
post-measurement gating\cite{diamanti06} is required for noise
rejection.  One possible drawback for the self-differencing detector
is that it does not allow tuning the gating frequency continuously
over a wide range, as the frequency is determined by the length
difference of two co-axial cables used. It is found that, within
$\pm0.5\%$ of the central gating frequency, the self-differencing
circuit cancels the capacitive responses well and no degradation in
photon counting performance was observed. Such a tuning range makes
it suitable for many applications, such as QKD, for which the clock
frequency has typically much better stability than $\pm0.5\%$.

In conclusion, we have demonstrated for the first time practical GHz
single photon detection at 1550~nm using a self-differencing InGaAs
APD.

\begin{figure} [c]
\begin{center}
\includegraphics[width=14cm]{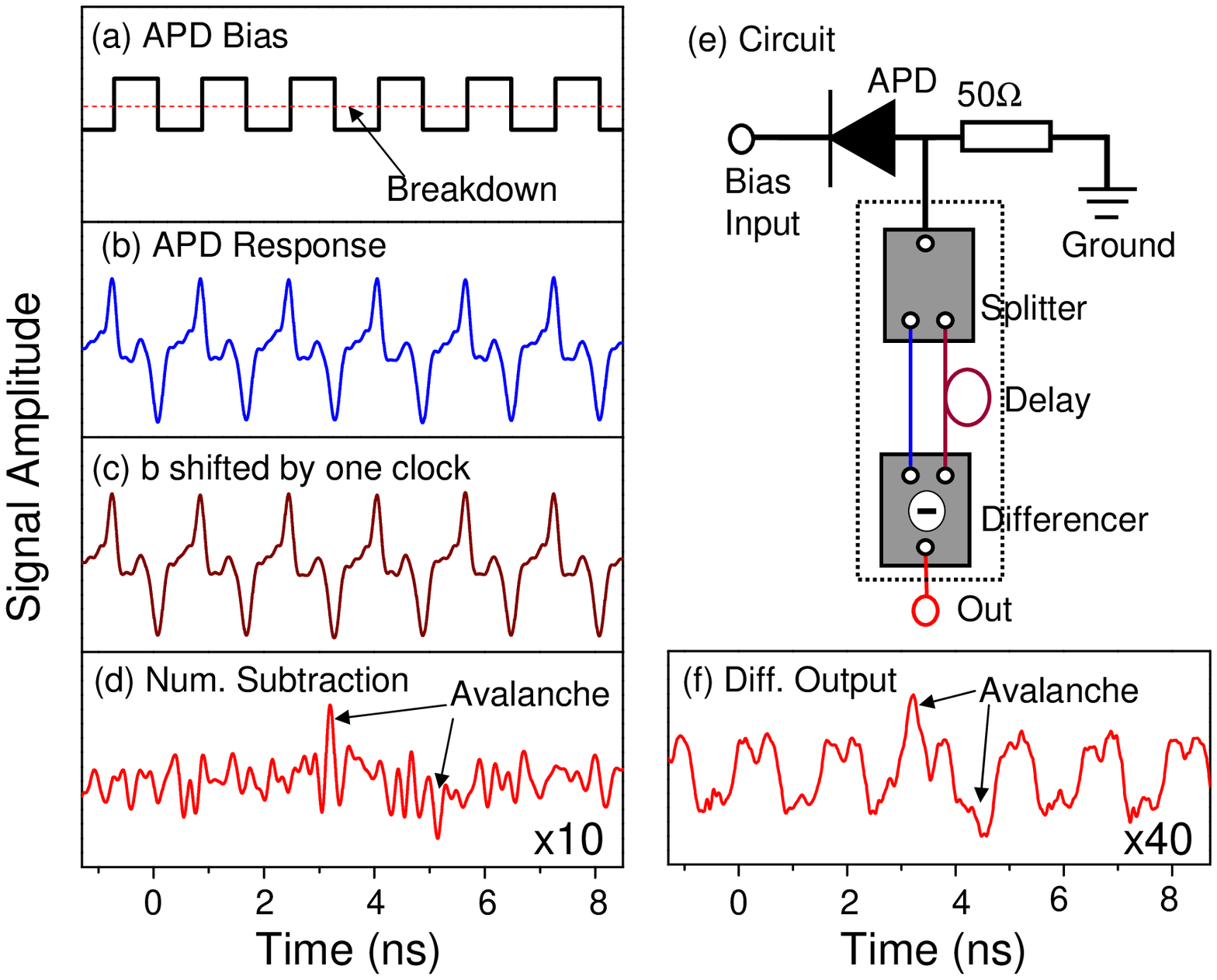}
\caption{(a) A series of biasing square wave gates (solid line)
applied to an APD; The dashed line indicates the APD breakdown
voltage. (b) An APD response to the square wave gates, recorded by
an oscilloscope; The APD here was gated with a continuous 0.62~GHz
square wave. Note that no avalanche is visible. (c) Same response as
(b), but shifted by a clock period; (d) Numerical subtraction
\textbf{b--c}, leaving the avalanche signal visible; Vertical scale
in (d) is scaled up by a factor of 10 as compared to (b) and (c) for
clarity. (e) An electrical circuit to realize the self differencing;
(f) Output of the self-differencer recorded by an oscilloscope.
Vertical scale here is scaled up by a factor of 40.}
\end{center}
\end{figure}

\begin{figure}
\begin{center}
\includegraphics[width=14cm]{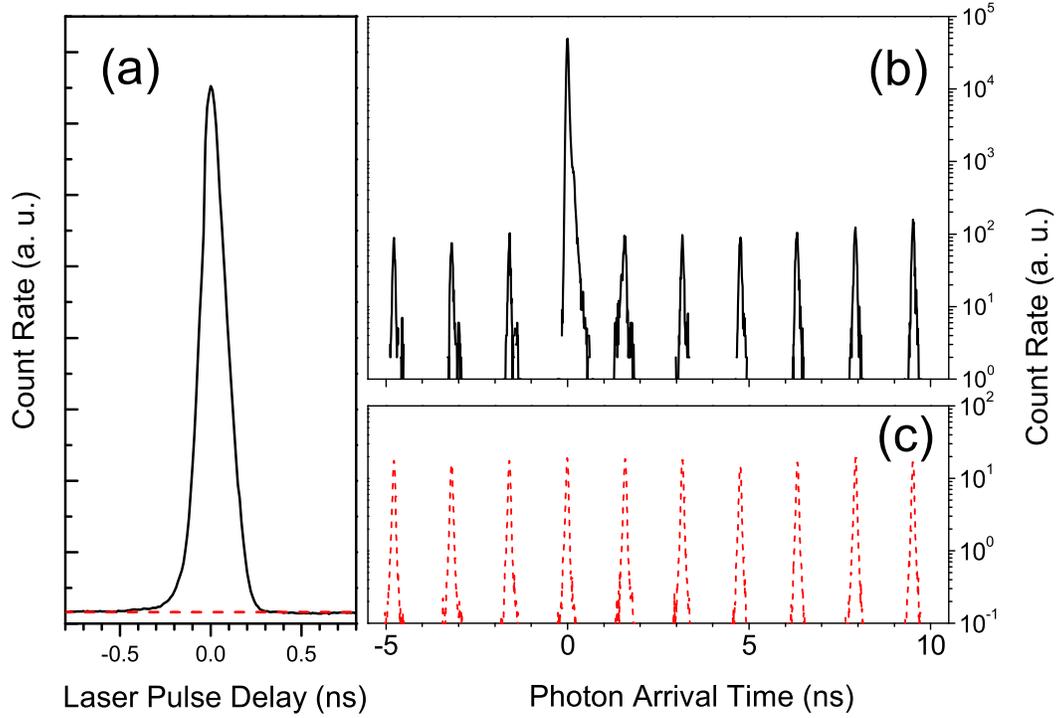}
\caption{(a) Photon count rate (solid line) as a function of the
laser pulse delay; The dashed line indicates the detector dark count
rate. (b) Time-resolved photon arrival time histogram under
illumination; (c) Time-resolved histogram of dark counts under no
illumination. The APD here was gated at 0.62~GHz.}
\end{center}
\end{figure}

\begin{figure}
\begin{center}
\includegraphics[width=14cm]{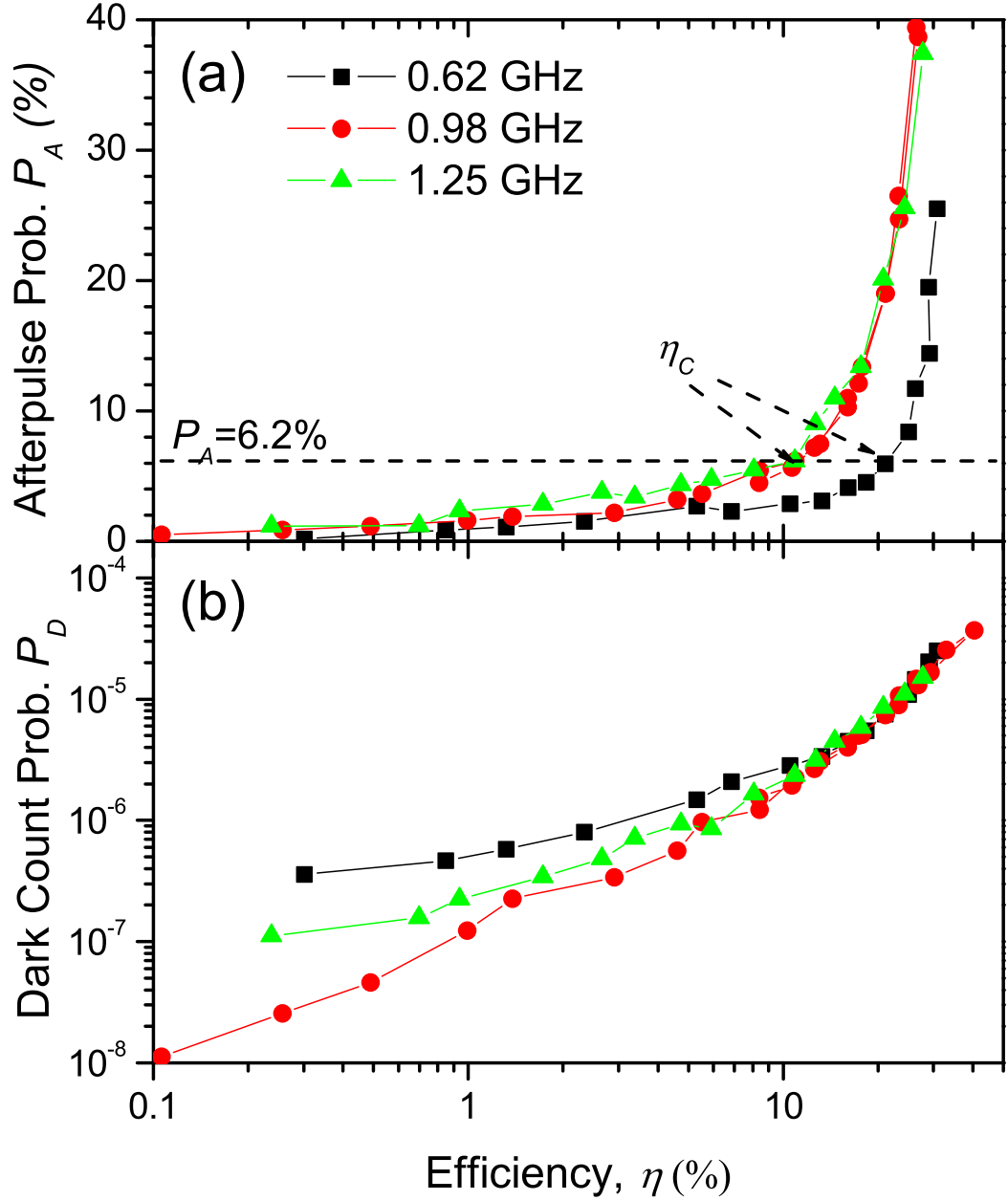}
\caption{(a) Detector afterpulse probability $P_{A}$, and (b), dark
count probability $P_D$ as a function of the net detection
efficiency $\eta$.}
\end{center}
\end{figure}

\begin{figure}
\begin{center}
\includegraphics[width=14cm]{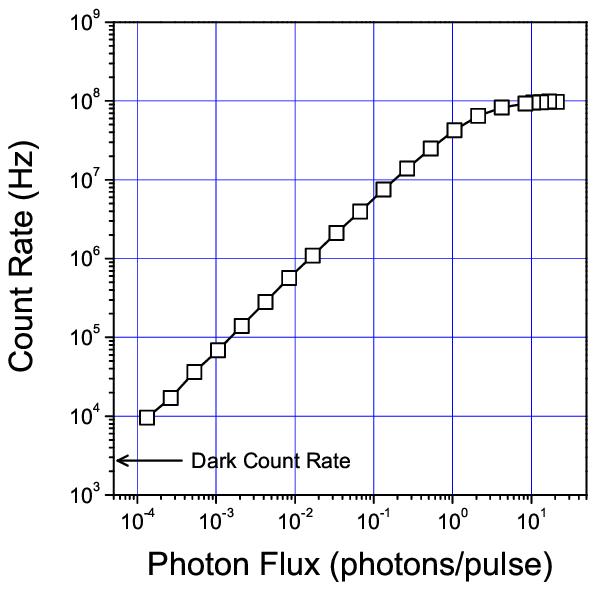}
\caption{Photon detection rate as a function of photon flux.
0.98~GHz gating was used for this measurement. The arrow indicated
the detector dark count rate of 2.8~kHz, which was subtracted from
the measured results.}
\end{center}
\end{figure}


\begin{thebibliography} {99}
\bibitem {perkinelmer} No maximum count rate of PPLN upconversion
detectors has been reported, and the value here is inferred from the
performance of Si-APD single photon detectors. See: Datasheet,
SPCM-AQR Single Photon Counting Module, PerkinElmer.
\end{thebibliography}
\end{document}